\documentclass[prl,twocolumn,superscriptaddress,showpacs,letterpaper]{revtex4}

\usepackage{amsmath,amsfonts}
\usepackage{graphicx}

\begin{document}


\title{Spin current and rectification in one-dimensional electronic systems}

\author{Bernd Braunecker}
\affiliation{Department of Physics, Brown University, Providence, RI 02912}
\affiliation{Department of Physics and Astronomy, University of Basel, 
             Klingelbergstrasse 82, CH-4056 Basel, Switzerland}

\author{D. E. Feldman}
\affiliation{Department of Physics, Brown University, Providence, RI 02912}

\date{\today}




\begin{abstract}
Spin and charge currents can be generated by an ac voltage through
a one-channel quantum wire with strong electron interactions in a 
static uniform magnetic field. 
In a certain range of low voltages, the spin current can grow as a 
negative power of the voltage bias as the voltage decreases.
The spin current expressed in units of $\hbar/2$ per second can become 
much larger than the charge current in units of the electron charge 
per second. The system requires neither spin-polarized particle
injection nor time-dependent magnetic fields.
\end{abstract}


\maketitle


The pioneering paper by Christen and B\"{u}ttiker \cite{Christen96} 
has stimulated much interest to rectification in quantum wires and 
other mesoscopic systems. Most attention was focused on the simplest 
case of Fermi liquids \cite{FLRect,magn}. Recently this research was 
extended to strongly interacting systems where Luttinger liquids 
are formed \cite{Feldman05,BB05b,magn-exp,magn-lat}.
One of the topics of current interest is the rectification effect 
in mesoscopic conductors in a magnetic field \cite{magn,magn-exp,magn-lat}.
In the presence of a magnetic field, both spin and charge currents 
can be generated. However, only charge currents have been studied so far. 
In this paper we consider the generation of a dc spin current by an 
ac voltage bias in a one-channel quantum wire.      
                                                                                
In recent years many approaches to the generation of spin currents in 
quantum wires were put forward. Typically both, spin and charge currents 
are generated and the spin current expressed in units of         
$\hbar/2$ per second is smaller than the electric current in units of $e$       
per second ($e$ is the electron charge). Such a situation naturally emerges
in partially polarized systems since each electron carries the charge $e$ and
its spin projection on the $z$-axis is $\pm\hbar/2$. 
A proposal how to obtain a spin current exceeding the charge current in 
a quantum wire was published by Sharma and Chamon \cite{Sharma,hall} who 
considered a Luttinger liquid in the presence of a time-dependent 
magnetic field in a region of the size of an electron wavelength. 
We show that the generation of a dc spin current exceeding the charge current 
is also possible without time-dependent magnetic fields
on the nanoscale. The spin current can be generated in a spatially asymmetric 
system in the presence of an ac bias.
Interestingly, in a certain interval of low voltages the dc spin current 
grows as a negative power of the ac voltage when the voltage decreases.   

\begin{figure}
\begin{center}
\includegraphics[width=\columnwidth]{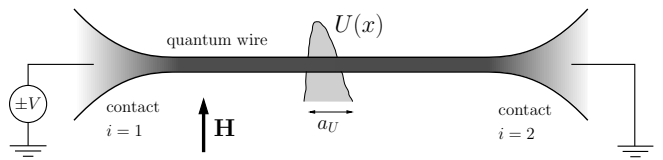}
\caption{
Sketch of the one-dimensional conductor connected to two electrodes
on both ends.
Currents are driven through a voltage bias $V$ that is applied on 
the left electrode while the right electrode is kept on ground.
The system is magnetized by the field $\mathbf{H}$. Electrons are 
backscattered off the asymmetric potential $U(x)$. $U(x)\ne 0$ in 
the region of size $a_U\sim 1/k_F$.
\label{fig:system}}
\end{center}
\end{figure}

The rectifying quantum wire 
is sketched in Fig.~\ref{fig:system}. 
It 
consists of a one-dimensional conductor with
a scatterer in the center of the system at $x=0$.
The scatterer creates an asymmetric potential $U(x)\ne U(-x)$.
The size of the scatterer $a_U\sim 1/k_F$ is of the order of the electron 
wavelength. The wire is placed in a uniform magnetic field $\mathbf{H}$.
The field defines the $S_z$ direction of the electron spins.
At its two ends, the wire is connected to nonmagnetic electrodes, labeled 
by $i=1,2$. The left electrode, $i=1$, is controlled by an ac voltage source,
while the right electrode, $i=2$, is kept on ground. 
The magnetic field $\mathbf{H}$ breaks the symmetry between the two 
orientations of the electron spin.
In a uniform wire this would not result in a net spin current since the 
conductances of the spin-up and -down channels would be the same, 
$e^2/h$ \cite{CleanCurrent}, and the spin currents of the spin-up and
-down electrons would be opposite. In the presence of a potential barrier 
such cancellation does not occur \cite{Hikihara05}. In a system with strong 
electron interaction, the spatial asymmetry of the wire 
leads to an asymmetric $I-V$ curve, $I(V)\ne I(-V)$ \cite{Feldman05}. 
Thus, an ac voltage bias generates spin and charge dc currents, $I^r_s$ and 
$I^r_c$. 
In this paper we focus on the low-frequency ac bias. We define the 
rectification current as the dc response to a low-frequency square voltage 
wave of amplitude $V$: $I_s^r(V)=[I_s(V)+I_s(-V)]/2$ and 
$I_c^r(V)=[I_c(V)+I_c(-V)]/2$.
The above dc currents express via the currents of spin-up and -down electrons:
$I_c^r=I_\uparrow^r+I_\downarrow^r$, 
$I_s^r=(\hbar/2e)[I_\uparrow^r-I_\downarrow^r]$. 
The spin current exceeds the charge current if the signs of  $I^r_{\uparrow}$ 
and $I^r_{\downarrow}$ are opposite.

The calculation of the rectification currents reduces to the calculation of
the contributions even in the voltage $V$ to the dc $I-V$ curves $I_s(V)$ 
and $I_c(V)$.
We assume that the Coulomb interaction between distant charges is screened 
by the gates. This will allow us to use the standard Tomonaga-Luttinger model 
with short range interactions \cite{Kane92}. Electric fields of external 
charges are also assumed to be screened.
Thus, the applied voltage reveals itself only as the difference
of the electrochemical potentials $E_1$ and $E_2$ of the
particles injected from the left and right reservoirs.
We assume that one lead is connected to the ground so
that its electrochemical potential $E_2=E_F$ is fixed. The
electrochemical potential of the second lead $E_1=E_F+eV$ is controlled by 
the voltage source.
Since the Tomonaga-Luttinger model captures only low-energy physics, we 
assume that $eV\ll E_F$, where $E_F$ is of the order of the bandwidth. 

Rectification occurs due to backscattering off the asymmetric potential 
$U(x)$. We will assume that the asymmetric potential is weak, $U(x)\ll E_F$.
This will enable us to use perturbation theory. As shown in 
Ref. \cite{Feldman05}, the leading contribution to the rectification effect 
emerges in the third order in $U(x)$. This fact allows for a simple 
explanation of our main prediction that the spin rectification current can 
significantly exceed the charge current. To illustrate the principle, let us 
consider the following toy problem: There is no uniform magnetic 
field $\mathbf{H}$ and no asymmetric potential $U(x)$. Instead, both 
right$\leftrightarrow$left and spin-up$\leftrightarrow$spin-down symmetries 
are broken by a weak coordinate-dependent magnetic field $B_z(x)\ne B_z(-x)$, 
which is localized in a small region of size $\sim 1/k_F$ (we do not include 
the components $B_{x,y}$ in the toy model). Let us also assume that the 
spin-up and -down electrons do not interact with the electrons of the opposite
spin. Then the system can be described as the combination of two 
spin-polarized one-channel wires with opposite spin-dependent potentials 
$\pm\mu B_z(x)$, where $\mu$ is the electron magnetic moment. According to 
Ref. \cite{Feldman05} an ac bias generates a rectification current in each 
of those two systems and the currents are proportional to the cubes of the 
potentials $(\pm \mu B_z)^3$.
Thus, $I^r_\uparrow=-I^r_\downarrow$. Hence, no net charge current 
$I^r=I^r_\uparrow+I^r_{\downarrow}$ is generated in the leading order. 
At the same time, there is a nonzero spin current in the third order in $B_z$.

Let us now calculate the currents in the presence of the asymmetric potential
$U(x)$ and the field $\mathbf{H}$. We assume that the magnetic field couples 
only to the electron spin and neglect the correction $-e \mathbf{A}/c$ to the 
momentum in the electron kinetic energy. Indeed, for a uniform field
one can choose $\mathbf{A}\sim y$, where the $y$-axis is orthogonal to the 
wire, and $y$ is small inside a narrow wire.
As shown in Ref. \cite{Hikihara05}, such a system allows a formulation within
the bosonization language and, in the absence of the asymmetric potential, 
can be described by a quadratic bosonic Hamiltonian
\begin{equation} \label{eq:H0}
	H_0 = 
	\sum_{\nu,\nu' = L,R}\sum_{\sigma,\sigma'=\uparrow,\downarrow}
	\int dx 
	\bigl(\partial_x \phi_{\nu\sigma}\bigr) 
	\mathcal{H}_{\nu\sigma,\nu'\sigma'}
	\bigr(\partial_x \phi_{\nu'\sigma'}\bigr),
\end{equation}
where $\sigma$ is the spin projection and $\nu=R,L$ labels the left and right 
moving electrons, which are related to the boson fields $\phi_{\nu\sigma}$ as 
$\psi^\dagger_{\nu\sigma}(x) \sim \eta^\dagger_{\nu\sigma}
\mathrm{e}^{\pm i(k_{F\nu\sigma} x + \phi_{\nu\sigma}(x))}$
with $\pm$ for $\nu = R,L$.
The operators $\eta^\dagger_{\nu\sigma}$ are the Klein factors adding a 
particle of type $(\nu,\sigma)$ to the system, and $k_{F\nu\sigma}/\pi$ is 
the density of $(\nu,\sigma)$ particles in the system.
The densities of the spin-up and -down electrons are different since 
the system is polarized by the external magnetic field.
The $4 \times 4$ matrix $\mathcal{H}$ describes the electron-electron
interactions. In the absence of spin-orbit interactions,
$L \leftrightarrow R$ parity is conserved and we can 
introduce the quantities $\phi_\sigma = \phi_{L\sigma}+\phi_{R\sigma}$
and $\Pi_\sigma = \phi_{L\sigma}-\phi_{R\sigma}$ such that the Hamiltonian
decouples into two terms depending on $\phi_\sigma$ and $\Pi_\sigma$
only.
In the absence of the external field, this 
Hamiltonian would further be diagonalized by the combinations 
$\phi_{c,s} \propto \phi_{\uparrow} \pm \phi_{\downarrow}$, and similarly for
$\Pi_{c,s}$, expressing the spin and charge separation.
This is here no longer the case. If we focus on the $\phi$ fields only
(as $\Pi$ will not appear in the operators describing backscattering off 
$U(x)$), the fields diagonalizing the Hamiltonian, $\tilde{\phi}_{c,s}$, have 
a more complicated linear relation to $\phi_{\uparrow,\downarrow}$, which 
we can write as
\begin{equation} \label{eq:transf}
	\begin{pmatrix} \phi_\uparrow \\ \phi_\downarrow \end{pmatrix}
	= 
	\begin{pmatrix} 
		\sqrt{g_c}[1+\alpha] &  \sqrt{g_s} [1+\beta] \\
		\sqrt{g_c}[1-\alpha] & -\sqrt{g_s} [1-\beta]
	\end{pmatrix}
	\begin{pmatrix} \tilde{\phi}_c \\ \tilde{\phi}_s \end{pmatrix},
\end{equation}
and which corresponds to the matrix $A$ of \cite{Hikihara05}.
The normalization has been chosen such that the propagator of the 
$\tilde{\phi}$ fields with respect to the Hamiltonian \eqref{eq:H0} evaluates 
to 
$\langle \tilde{\phi}_{c,s}(t_1) \tilde{\phi}_{c,s}(t_2) \rangle 
= - 2 \ln(i(t_1-t_2)+\delta)$, where $\delta > 0$ is an infinitesimal quantity.
For non-interacting electrons without a magnetic field,
$g_c=g_s = 1/2$. $g_c < 1/2$ ($> 1/2$) for repulsive (attractive) interactions. 
In the absence of the magnetic field, terms in \eqref{eq:H0} in the form of 
$\exp(\pm 2i\sqrt{g_s}\phi_s)$ may become relevant and open a spin gap
for $g_s < 1/2$. 
In our model they can be neglected since they are suppressed by the rapidly 
oscillating factors $\exp(\pm2i[k_{F\uparrow}-k_{F\downarrow}]x)$.
It is convenient to model the leads as the regions near the right 
and left ends of the wire without electron interaction \cite{CleanCurrent}.

Backscattering off the impurity potential $U(x)$ is described by the following 
contribution to the Hamiltonian $H = H_0 +H'$ \cite{Kane92}:
\begin{equation} \label{eq:backscattering}
	H' = \sum_{n_\uparrow,n_\downarrow} 
	U(n_\uparrow,n_\downarrow)
	\mathrm{e}^{i n_\uparrow \phi_{\uparrow}(0) + i n_\downarrow \phi_{\downarrow}(0)},
\end{equation}
where the fields are evaluated at position $x=0$ and 
$U(n_\uparrow,n_\downarrow)=U^* (-n_\uparrow,-n_\downarrow)$
since the Hamiltonian is Hermitian. The fields $\Pi$ do not enter the above 
equation due to the conservation  of the electric charge and the 
$z$-projection of the spin.
The Klein factors are not written because they drop out in the perturbative 
expansion.
$U(n_\uparrow,n_\downarrow)$ are the amplitudes of backscattering of 
$n_\uparrow$ spin-up and $n_\downarrow$ spin-down particles
with $n_\sigma > 0$ for $L \to R$ and $n_\sigma <0$ for $R\to L$ scattering. 
$U(n_\uparrow,n_\downarrow)$  can be estimated as 
$U(n_\uparrow,n_\downarrow) \sim \int dx U(x) 
\mathrm{e}^{i n_\uparrow 2k_{F\uparrow}x+i n_\downarrow 2k_{F\downarrow}x}
\sim U/ k_F$,
where $U$ is the maximum of $U(x)$ (cf. \cite{Kane92}). 
In the case of a symmetric potential $U(x)=U(-x)$ the coefficients 
$U(n_\uparrow,n_\downarrow)$ are real.

The spin and charge current can be expressed as
\begin{equation}
	I_{s,c} = L_{s,c}^1 + R_{s,c}^1=L_{s,c}^2+R_{s,c}^2,
\end{equation} 
where $L_{s,c}^i$ and $R_{s,c}^i$ denote the current of the left-
and right-movers near electrode $i$, respectively.
For a clean system ($U(x)=0$), the currents $R_{s,c}^1=R_{s,c}^2$, 
$L_{s,c}^1=L_{s,c}^2$ and $I_c = 2e^2V/h$, $I_s=0$ \cite{CleanCurrent}.
With backscattering off $U(x)$, particles are transferred between $L$ 
and $R$ in the wire, and hence $R_{s}^2=R_{s}^1+dS_R/dt$, 
$R_c^2=R_c^1+dQ_R/dt$, where $Q_R$ and $S_R$ denote the total charge 
and the $z$-projection of the spin of the right-moving electrons
\cite{Feldman03}.
The currents $L_{s,c}^2$ and $R_{s,c}^1$ are determined by 
the leads (i.e. the regions without electron interaction in our 
model \cite{CleanCurrent}) and remain the same as in the absence
of the asymmetric potential. Thus, the spin and charge current can 
be represented as $I_c=2e^2V/h+I_c^{bs}$ and $I_s=I_s^{bs}$, where 
the backscattering current operators are \cite{Feldman05,BB05b,Kane92}
\begin{align} 
\label{eq:I_c}
	\hat{I}_c^{bs} 
	&= {d\hat Q_R}/{dt}=ie[H,\hat{Q}_R]/\hbar  \nonumber \\
	&=\frac{-ie}{\hbar}
	\sum_{n_\uparrow,n_\downarrow}(n_\uparrow+n_\downarrow)
	U(n_\uparrow,n_\downarrow)
	\mathrm{e}^{i n_\uparrow \phi_{\uparrow}(0) + i n_\downarrow \phi_{\downarrow}(0)},
\\
\label{eq:I_s}
	\hat{I}_s^{bs}  
	&={d\hat{S}_R}/{dt}= 
	-\frac{i}{2}\sum_{n_\uparrow,n_\downarrow}(n_\uparrow-n_\downarrow)
	U(n_\uparrow,n_\downarrow)
	\mathrm{e}^{i n_\uparrow \phi_{\uparrow} + i n_\downarrow \phi_{\downarrow}}. 
\end{align}
The calculation of the rectification currents reduces to the calculation of 
the currents \eqref{eq:I_c}, \eqref{eq:I_s} at two opposite values of the 
dc voltage.

To find the backscattered current we use the Keldysh technique
\cite{Keldysh}. We assume that at $t=-\infty$ there is no backscattering
in the Hamiltonian ($U(x)=0$), and then the backscattering is
gradually turned on. Thus, at $t=-\infty$, the numbers $N_L$ and $N_R$ of the 
left- and right-moving electrons conserve separately: 
The system can be described by a partition function with two chemical 
potentials $E_1=E_F+eV$ and $E_2=E_F$ conjugated with the
particle numbers $N_R$ and $N_L$. This initial state determines
the bare Keldysh Green functions. 

We will consider only the zero temperature limit. It is convenient
to switch \cite{Feldman03} to the interaction representation 
$H_0\rightarrow H_0-E_1 N_R-E_2 N_L$. 
This transformation induces a time dependence in the electron 
creation and annihilation operators.
As the result each exponent in Eq.~\eqref{eq:backscattering} 
is multiplied by $\exp(ieVt[n_\uparrow+n_\downarrow]/\hbar)$.

In the Keldysh formulation \cite{Keldysh} the backscattering currents 
\eqref{eq:I_c}, \eqref{eq:I_s} are evaluated as
\begin{equation} \label{eq:I_Keldysh}
	I_{c,s}^{bs}
	= \langle 0 | \mathcal{S}(-\infty,0) 
	  \hat{I}_{c,s}^{bs} \mathcal{S}(0,-\infty) | 0 \rangle,
\end{equation}
where $|0\rangle$ is the ground state for the Hamiltonian $H_0$, 
Eq. \eqref{eq:H0}, and $\mathcal{S}(t,t')$ the evolution operator for 
$H'$ from $t'$ to $t$ in the interaction representation with respect 
to $H_0$. The result of this calculation depends on the elements of 
the matrix \eqref{eq:transf}, which describe the low-energy degrees of 
freedom and depend on the microscopic details. Several regimes
are possible \cite{BB05b} at different values of the parameters 
$g_s>0$, $g_c>0$, $\alpha$ and $\beta$. In this paper
we focus on one particular regime:
\begin{equation} \label{eq:ineq}
	1\gg g_c>g_s\gg g_c-g_s\gg \alpha,\beta.
\end{equation}
We will see that in such a regime the spin current can be much greater 
than the charge current. 
We will also assume that $[g_c\alpha+g_s\beta]>0$ but our results do 
not depend significantly on this assumption.
The behavior which we find in
the regime (8) persists for $g_c,g_s,\alpha$ and $\beta$ smaller than some
constants of the order of unity. In this paper we focus on the limit in
which the current can be calculated analytically. A numerical investigation
of the boundary of the region where the spin current exceeds the charge
current is beyond the scope of the present paper.

The currents \eqref{eq:I_Keldysh} can be estimated using a 
renormalization group procedure \cite{Kane92}. 
As we change the energy scale $E$, the backscattering amplitudes 
$U(n_\uparrow,n_\downarrow)$ scale as $U(n,m;E)\sim E^{z(n,m)}$, 
where the scaling dimensions are
\begin{eqnarray} \label{eq:z}
	z(n,m) = n^2 [ g_c (1+\alpha)^2 + g_s (1+\beta)^2 ]
	       + m^2 [ g_c (1-\alpha)^2 & & \nonumber\\
	              +g_s (1-\beta)^2 ]
		   + 2 nm [ g_c (1-\alpha^2) - g_s (1-\beta^2) ] - 1. & &
\end{eqnarray}
The renormalization group stops on the scale of the order $E\sim eV$. 
At this scale the backscattering current can be represented as 
$I_{c,s}^{bs} = V r_{c,s}(V)$, where the effective reflection coefficient 
$r_{c,s}(V)$ \cite{Feldman05,Kane92} is given by the sum of contributions 
of the form 
$(\mathrm{const})U(n_1,m_1)U(n_2,m_2)\dots U(n_p,m_p)
V^{\sum_{k=1}^p z(n_k,m_k)}$.
Such a perturbative expansion can be used as long as $U$ is small. 
In the regime \eqref{eq:ineq} it is sufficient
to require that $U(n,m)<(eV)^{1+\delta}/E_F^\delta$, where $\delta>g_c$.
The contributions to the backscattering current of the $n$th order 
in $U$ scale as 
$I_{c,s}^{(n)}\sim U^n V^{1-n+O(g_c)}$. Hence, they exceed the contributions
of the $(n+1)$th order for $U(n,m)<(eV)^{1+\delta}/E_F^\delta$. 
The leading non-zero contribution to the backscattered current, therefore,
emerges in the second order. The \emph{rectification current}, however,
is dominated by a third order contribution for $V^{1+\delta}>U\gg V^2/E_F$. 

Indeed, the second order contributions to the charge current were computed 
in Ref. \cite{Kane92}. The spin current can be found
in exactly the same way. The result is 
\begin{equation} \label{eq:2order}
	I_{c,s}^{(2)bs}(V)
	\sim
	\sum (\mathrm{const}) 
	|U(n,m)|^2|V|^{2z(n,m)+1}\mathrm{sign} V.
\end{equation}
If the $U(n,m)$ were independent of the voltage, the above current would be 
an odd function of the bias and hence would not contribute to the 
rectification current. The backscattering amplitudes depend \cite{Kane92} on 
the charge densities $k_{F\nu\sigma}$ though, which in turn depend on the 
voltage in our model \cite{Feldman05}. The voltage-dependent corrections to 
the amplitudes are linear in the voltage at low bias. Hence, the second order 
contributions to the rectification currents scale as $U^2|V|^{2z(n,m)+2}$. 
The additional factor of $V$ makes the second order contribution smaller than 
the leading third order contribution \eqref{eq:spin} at $U\gg V^2/E_F$. 
Note that the second order contribution to the rectification current is 
nonzero even for a symmetric potential $U(x)$. The leading third order 
contribution emerges solely due to the asymmetry of the scatterer.

It is easy to find the most relevant backscattering operators in the 
renormalization group sense using Eqs. \eqref{eq:ineq} and \eqref{eq:z}.
The most relevant operator is $U(1,0)$, the second most relevant 
$U(0,-1)$, and the third most relevant $U(-1,1)$. The leading non-zero 
third order contributions to the spin and charge currents come from 
the product of the above three operators in the Keldysh perturbation 
theory. This leads to
\begin{equation} \label{eq:scaling}
	I^{bs}_{c,s}\sim U^3 V^{g_c(2+6\alpha^2)+g_s(6+2\beta^2)-2}.
\end{equation}
This contribution dominates at $(eV)^2/E_F\ll U \ll (eV)$. Interestingly, 
the current \eqref{eq:scaling} grows as the voltage decreases. 

Does the current \eqref{eq:scaling} contribute to the rectification current? 
In general, \eqref{eq:scaling} is the sum of odd and even functions of the 
voltage and only the even part is important for us. One might naively expect 
that such a contribution has the same order of magnitude for the spin and 
charge currents. A direct calculation shows, however, that this is not the 
case and the spin rectification current is much greater than the charge 
rectification current.

In order to calculate the prefactors in the right hand side of 
Eq. \eqref{eq:scaling} one has to employ the Keldysh formalism.
The third order Keldysh contribution reduces to the integral of 
$P(t_1,t_2,t_3)=\langle T_c\exp(i\phi_\uparrow(t_1)+ieVt_1/\hbar)
\exp(-i\phi_\downarrow(t_2)-ieVt_2/\hbar)
\exp(i[-\phi_\uparrow(t_3)+\phi_\downarrow(t_3)])\rangle$
over $(t_1-t_3)$ and $(t_2-t_3)$, where $T_c$ denotes time ordering along 
the Keldysh contour $-\infty \to 0 \to -\infty$ and the angular 
brackets denote the average with 
respect to the ground state of the non-interacting Hamiltonian \eqref{eq:H0}. 
For the purpose of comparing the prefactors in Eq. \eqref{eq:scaling} 
for $I^{bs}_{c}$ and $I_{s}^{bs}$ in the region of parameters \eqref{eq:ineq},
it is sufficient to perform the calculation in the limit $1\gg g_c=g_s$, 
$\alpha=\beta=0$, which corresponds to the boundary
of the region \eqref{eq:ineq}. A small change of the parameters 
$g_c,g_s,\alpha$ and $\beta$ will not significantly affect 
the prefactors (as well as the exponent in Eq. \eqref{eq:scaling}). 
In the above limit the calculation considerably simplifies and can 
be performed analytically. The correlation function $P(t_1,t_2,t_3)$ 
factorizes into the product of a function of $(t_1-t_3)$ and a 
function of $(t_2-t_3)$. One finds
\begin{align} \label{eq:charge}
	I^{bs}_c
	= &-\frac{8e\tau_c^2}{\hbar^3} \mathrm{sign}(eV)
	\left| \frac{eV\tau_c}{\hbar}\right|^{8g-2} 
	\Gamma^2(1-4g)\sin(4\pi g) 
  \nonumber \\
	&\times [1-\cos(4\pi g)] \mathrm{Re}[U(1,0)U(-1,1)U(0,-1)],
\\
\label{eq:spin}
	I^{bs}_s
	= &-\frac{4\tau_c^2}{\hbar^2}
	\left|\frac{eV\tau_c}{\hbar}\right|^{8g-2} 
	\Gamma^2(1-4g) 
  \nonumber \\
	&\times\sin^2(4\pi g)
	\mathrm{Im}[U(1,0)U(-1,1)U(0,-1)],
\end{align}
where $g=g_c=g_s$ and $\tau_c\sim \hbar/E_F$ is the ultraviolet cutoff time. 
The above results apply at $g<1/4$, i.e., when the Keldysh integral converges. 
The charge current \eqref{eq:charge} is an odd function of the voltage and 
hence does not contribute to the rectification effect. The spin 
current \eqref{eq:spin} is an even function and hence determines the spin 
rectification current. It is non-zero if 
$\mathrm{Im}[U(1,0)U(-1,1)U(0,-1)]\ne 0$,
which is satisfied for asymmetric potentials.
Thus, we expect that in the region \eqref{eq:ineq}, the spin rectification 
current exceeds the charge rectification current
in an appropriate interval of low voltages $\sqrt{UE_F}\gg eV \gg U$.

\begin{figure}
\begin{center}
\includegraphics[width=\columnwidth]{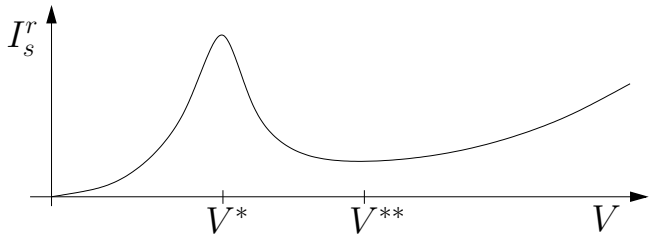}
\caption{
Qualitative representation of the spin rectification current.
The spin current exceeds the charge current and follows a power-law
dependence on the voltage with a negative exponent
in the interval of voltages $V^{*}<V<V^{**}$.
\label{fig:currents}}
\end{center}
\end{figure}

The voltage dependence of the spin rectification current is illustrated 
in Fig. \ref{fig:currents}. The expression \eqref{eq:scaling}
describes the current in the voltage interval $V^{**}>V>V^{*}$. 
In this interval the current \emph{increases} as the voltage decreases. 
At lower voltages the perturbation theory breaks down. The current must 
decrease as the voltage decreases below $V^{*}$ and eventually reach 0 
at $V=0$. At higher voltages, $E_F\gg eV\gg eV^{**}$, the second order 
rectification current \eqref{eq:2order} dominates. The leading second 
order contribution $I^r_{s}\sim |U(1,0)|^2V^{2z(1,0)+2}$ grows as the 
voltage increases. The charge rectification current has the same order 
of magnitude as the spin current.
The Tomonaga-Luttinger model cannot be used for the highest voltage 
region $eV\sim E_F$.


In conclusion, we have shown that rectification in magnetized quantum 
wires can lead to a spin current that largely exceeds the charge current. 
The effect is solely due to the properties of the wire and does not require
spin polarized injection as from magnetic electrodes. 
The currents are driven by the voltage source only. In an interval of low 
voltages the spin current grows as the voltage decreases.


We thank J. B. Marston for many helpful discussions.
This work was supported in part by the NSF under grant numbers DMR-0213818, 
DMR-0544116, and PHY99-07949, and by Salomon Research Award. 
D.E.F. acknowledges the hospitality of the Aspen Center for Physics,
of the MPI Dresden, and of the KITP Santa Barbara 
where this work was completed.


\begin{thebibliography}{7}

\bibitem{Christen96}
T. Christen and M. B\"{u}ttiker,
	Europhys. Lett. \textbf{35}, 523 (1996).

\bibitem{FLRect}
P. Reimann, M. Grifoni, and P. H\"{a}nggi,
	Phys. Rev. Lett. \textbf{79}, 10 (1997);
J. Lehmann, S. Kohler, P. H\"{a}nggi, and A.~Nitzau,
	\textit{ibid.} \textbf{88}, 228305 (2002);
S. Scheidl and V. M. Vinokur,
	Phys. Rev. B \textbf{65}, 195305 (2002).

\bibitem{magn}
D. S\'{a}nchez and M. B\"{u}ttiker,
	Phys. Rev. Lett. \textbf{93}, 106802 (2004);
B. Spivak and A. Zyuzin,
	\textit{ibid.} \textbf{93}, 225801 (2004).

\bibitem{Feldman05}
D.~E. Feldman, S. Scheidl, and V. Vinokur,
	Phys. Rev. Lett. \textbf{94}, 186809 (2005).

\bibitem{BB05b}
B. Braunecker, D.~E. Feldman, and J.~B. Marston,
	Phys. Rev. B \textbf{72}, 125311 (2005).

\bibitem{magn-exp}
V. Krstic, S. Roth, M. Burghard, K. Kern, and G.~L.~J.~A. Rikken, 
	J. Chem. Phys. \textbf{117}, 11315 (2002);
J. Wei, M. Shimogawa, Z. Wang, I. Radu, R. Dormaier, and D.~H. Cobden,
	 Phys. Rev. Lett. \textbf{95}, 256601 (2005).

\bibitem{magn-lat} 
A. De Martino, R. Egger, and A.~M. Tsvelik,
	Phys. Rev. Lett. \textbf{97}, 076402 (2006).

\bibitem{Sharma}
P.~Sharma and C.~Chamon, 
	Phys. Rev. Lett. \textbf{87}, 096401 (2001);
P.~Sharma,
	Science \textbf{307}, 531 (2005).

\bibitem{hall}
Spin current without charge current was also predicted in the 
context of the quantum Hall effect in graphene \cite{Abanin06}.

\bibitem{Abanin06}
D. A. Abanin, P. A. Lee, and L. S. Levitov, 
	Phys. Rev. Lett. \textbf{96}, 176803 (2006).

\bibitem{CleanCurrent}
D.~L. Maslov and M. Stone, 
	Phys. Rev. B \textbf{52}, R5539 (1995);
V.~V. Ponomarenko, 
	\textit{ibid.},  R8666 (1995);
I. Safi and H.~J. Schulz, 
	\textit{ibid.}, R17040 (1995).

\bibitem{Hikihara05}
T. Hikihara, A. Furusaki, and K.~A. Matveev,
	Phys. Rev. B \textbf{72}, 035301 (2005).

\bibitem{Kane92}
C.~L. Kane and M.~P.~A. Fisher,
	Phys. Rev. B \textbf{46}, 15233 (1992);
A. Furusaki and N. Nagaosa,
	\textit{ibid.} \textbf{47}, 4631 (1992).

\bibitem{Feldman03}
D. E. Feldman and Y. Gefen,
	Phys. Rev. B \textbf{67}, 115337 (2003).

\bibitem{Keldysh}
L.~V. Keldysh,
	Sov. Phys. JETP \textbf{20}, 1018 (1965);
J. Rammer and H. Smith,
	Rev. Mod. Phys. \textbf{58}, 323 (1986).

\end{thebibliography}
\end{document}